\documentstyle[aps,prb,epsf,eqsecnum]{revtex}
\newcommand{\be}{\begin{equation}}
\newcommand{\ee}{\end{equation}}
\newcommand{\ba}{\begin{eqnarray}}
\newcommand{\ea}{\end{eqnarray}}
\newcommand{\cdag}{c^{\dag}}

\begin{document}
\draft
\title{Green's Function Approach to the Edge Spectral Density}
\author{ J. H. Han}
\address{Department of Physics, Box 351560, University of Washington,
Seattle, Washington 98195}

\maketitle
\begin{abstract}
It is shown that the conventional many-body techniques to calculate the
Green's functions can be applied to the wide, compressible edge of a
quantum Hall bar. The only {\it ansatz} we need is the existence of
stable density modes that yields a simple equation of motion of the
density operators. We derive the spectral density at a finite temperature 
and show how the tunneling characteristics of a sharp edge can be
deduced as a limiting case.
\end{abstract}
%\narrowtext
\pacs{PACS numbers: 73.40.Hm, 73.20.Mf}

\section{Introduction}
Green's functions are  a powerful tool for studying many-body
phenomena. Since it is generally impossible to solve the many-body problem
exactly, one usually relies on a set of approximation rules that will 
yield the
calculation tractable. In one version, the difficulty is expressed in the 
fact that the solution of the $N$-particle Green's function requires
the knowledge of $N+1$-particle function and so on {\it ad infinitum.} 
In the RPA theory of the collective oscillation,\cite{bohm,pines}
the density operator obeys a Heisenberg equation of motion 
like

\be
 -i\dot{\rho}(k)=[H,\rho(k)]=\omega(k)\rho(k),
\label{eq:RPAeq}
\ee
where $\omega(k)$ is generally linear in $k$. Bohm and Pines\cite{bohm}
pointed out that Tomonaga's idea of quantizing the density 
operators\cite{tomonaga} 
are identical to their RPA theory specialized to one dimension. Much 
later it was realized  by Everts and Schulz\cite{everts} that, by 
using the above equation for the density, the Green's function
hierarchy terminates at the two-particle level in one dimension, 
and one can solve for the single-particle 
spectral density exactly. They showed their results are equivalent to
those previously obtained by other methods such as bosonization.
\cite{voit}

The advantage of Everts' approach is that, since it is based on a more
conventional technique, one can generalize it to other situations
where the rigorous proof of fermion-boson equivalence may not be 
available. A compressible edge of the quantum Hall liquid
\cite{csg}
is one such example of nearly one-dimensional
systems where it is not entirely certain whether a strict bosonization theory
is a valid scheme to try. Conti and Vignale\cite{conti} and Han and 
Thouless\cite{han} have independently produced schemes general 
enough to accommodate the wide edge dynamics that eventually reduce to the
bosonization result for the sharp edge limit. 
The latter approach is based on the idea of treating the density 
fluctuation as a hydrodynamic excitation and subsequently quantizing the 
classical action. The results are identical to those obtained from the 
independent boson model of Conti and Vignale. The validity of 
either approaches was not to be answered, however, 
unless one had a more general starting point. The Green's function 
approach, to be explained in this paper, provides such a starting 
point. It is shown that the  validity of the bosonization is closely 
tied to the random phase approximation, namely that one has a
collection of {\it stable} density modes obeying equations like
Eq.\ (\ref{eq:RPAeq}).

The starting point of our analysis is the second-quantized Hamiltonian
in the lowest Landau level. Section 2 outlines the basic formalism and 
derives the equation of motion for the density operator in the random 
phase approximation. The analysis of previous theories\cite{conti,han} 
assumes the edge is an equipotential surface as the classical 
solution\cite{csg} suggests. In section 3, we argue that the 
equipotential state is obtained as a result of the 
coarse-graining of the microscopic state.
In section 4, it is shown how the electronic spectral properties can be 
derived using 
techniques entirely analogous to those explored by Everts and Schulz. The 
conventional bosonization results are recovered as a special limit. For 
more than one density mode present, the spectral density is a product of 
the spectral densities associated with the individual density mode.
The electron tunneling current between two edges are derived in a closed 
form in section 5. The conclusion follows in section 6.

\section{Second Quantization}
Our starting point for the microscopic analysis of the edge dynamics is the
Hamiltonian

\be
 H = {e^2\over{2\kappa}}\int {1\over |r-r'|}
\psi^{\dag}(r)\psi^{\dag}(r')\psi(r')\psi(r) d^{2}rd^{2}r'
+\int V_{i}(r)\psi^{\dag}(r)\psi(r)d^{2}r.
\label{eq:operatorH}\ee
which consists of the Coulomb interaction and the external 
potential. There is no kinetic energy term as we are working exclusively 
within the lowest Landau level. We can write out the various operators in 
the basis of wavefunctions in the lowest Landau level. 
We will use $L$ for the length (perimeter) of the 
Hall bar (droplet).

\be
\psi(r)=\pi^{-1/4}L^{-1/2}\sum_{k}e^{ikx}
e^{-{1\over2}(y-k)^{2}}c_k.
\ee
When this expression is substituted in the Hamiltonian, Eq.\ 
(\ref{eq:operatorH}), we obtain the second-quantized version of the 
Hamiltonian in momentum space

\ba
      H &=& H_{C}+H_i, \nonumber \\
     H_{C}& = &  {1\over2}\sum_{k_1 k_2 q}\!\!V_{C}(q,k_1\!-\!k_2)
        \cdag_{k_1\!+\!q/2}\cdag_{k_2\!-\!q/2}
         c_{k_2\!+\!q/2}c_{k_1\!-\!q/2}, \nonumber \\
     H_{i}& = & \sum_{kq}V_{i}(q,k)\cdag_{k\!+\!q/2}c_{k\!-\!q/2}.
\ea
We have used the electronic basis to expand the Hamiltonian, and 
consequently our theory may be more suitable to describe the 
integer edge. There is no known way one can arrive at 
something like a Laughlin state from the above type of Hamiltonian and 
working one's way up through diagrammatic theory. With the wide edge, 
our theory may still be valid a sufficient distance away from the 
fractional bulk where fractional quasiparticle states 
are unlikely to exist, but is less certain to work close to the 
bulk where a small decrease in density from the bulk value may be better 
described as a dilute set of quasiholes.
We will adopt the above Hamiltonian as a starting point, if 
for lack of a better one, and see if something sensible can come out of 
it. Quite surprisingly, the electronic properties of the 
fractional edge such as spectral density and the electron tunneling 
current
follow very straightforwardly from the above Hamiltonian.  

The definition of the Coulomb and the external potential in the momentum 
representation are the following:

\ba
 V_{C}(q,k) & = & {{2e^2}\over{\sqrt{2\pi}\kappa L}}e^{-{q^2}/2}
              \!\int_{-\infty}^{\infty}dy
        K_{0}(|qk\!+\!qy|)e^{-{y^2}/2}, \nonumber \\
 V_{i}(q,k) & = & {1\over{\sqrt{\pi}L}}
  e^{-q^{2}/4}\int V_{i}(x,y)e^{-iqx-(y-k)^{2}}dxdy.
\label{eq:operator}
\ea
The kernel is the modified Bessel function $K_{0}$. 
For the external potential which depends on $y$ only, we obtain a simpler 
expression, 

\be
V_{i}(k)\!=\!V_{i}(0,k)\!=\!\pi^{-1/2}\int V_{i}(y)e^{-(y-k)^{2}}dy.
\ee
The confining potential of a Hall bar with the translation symmetry
in the $x$ direction would be such 
a case. In this paper, we will restrict ourselves to this form of the 
external potential.

\begin{figure}[ht]
\centering
\leavevmode
\epsfxsize=9cm\epsfysize=11cm\epsfbox{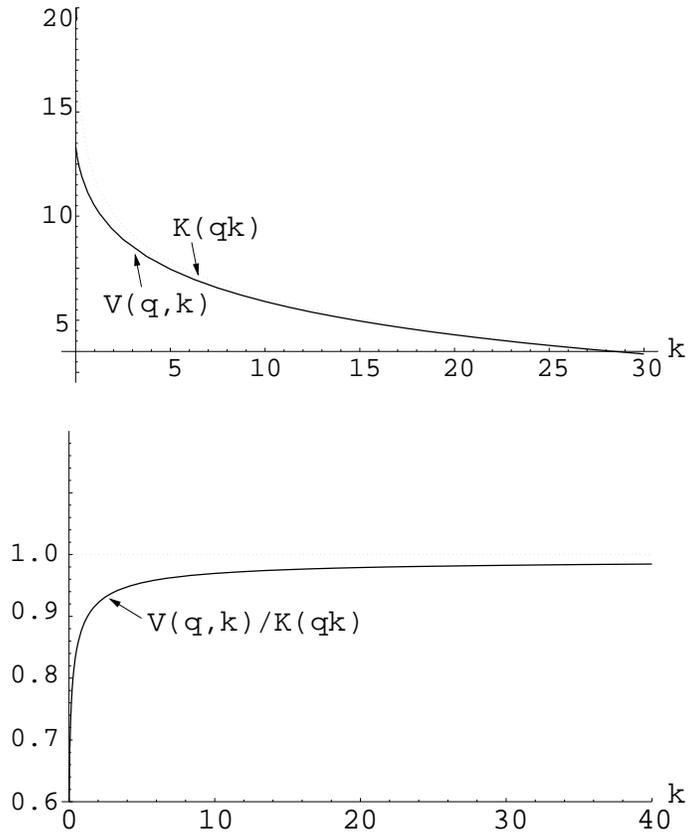}
\caption[Plot of the Classical and Quantum-Mechanical Kernel]
{Comparison of the classical kernel
(denoted as $K(qk)$) and its quantum counterpart, $V(q,k)$, for
$q=10^{-2}$. For $qk\approx 0.3$ the two curves are already
indistinguishable. In the bottom plot, the ratio
$V(q,k)/K(qk)$ is shown. The
classical kernel is always
larger, but the difference quickly diminishes as $k$ grows.}  
\label{kernel}
\end{figure}

Note that the integral for $V_{C}(q,k)$ is the convolution of the 
modified Bessel function with a gaussian of width $l_{0}$, which we have 
taken to be unity. For the limit of a very strong magnetic field, the 
integral is approximately equal to
$\sqrt{2\pi}K_{0}(|qk|)$. In the classical treatment of Aleiner and 
Glazman\cite{aleiner}, the kernel was 
simply this modified Bessel function. The classical theory does not 
explicitly take the size of the electron wavefunction into account which 
in our case is the magnetic length. When we do,
it turns out the relevant quantities are the convolution of the classical 
quantities with an envelope function of the width of the magnetic length as 
evidenced in Eq.\ (\ref{eq:operator}).
For clarity, we have plotted  $V(q,k)$ and 
$\sqrt{2\pi}K_{0}(|qk|)$ for $q=10^{-2}$ in Fig.~\ref{kernel}. 
The difference is minute in almost all regions of $k$ except 
very close to the origin. At $k=0$, the classical kernel diverges 
logarithmically, while $V_{C}(q,0)\!=\! 
(e^{2}/\kappa L)e^{-q^{2}/4}K_{0}(q^{2}/4)$ leading to a finite interaction 
energy even at zero distance.

We will work out the time development of
$c^{\dag}_{k\!+\!q/2}c_{k\!-\!q/2}$, since it is this type of operator that 
gives rise to the density fluctuation.
The commutator with the Coulomb part of the Hamiltonian gives

\ba
[H_{C},\cdag_{k+q/2}c_{k-q/2}]& = &
\sum_{pk'}V_{C}(p,k\!-\!k'\!+\!{1\over2}(q\!+\!p))
\cdag_{k\!+\!p\!+\!q/2}
\cdag_{k'\!-\!p/2}c_{k'\!+\!p/2}c_{k\!-\!q/2}
\nonumber \\
 & - &
\sum_{pk'}V_{C}(p,k\!-\!k'\!-\!{1\over2}(q+p))
\cdag_{k\!+\!q/2}\cdag_{k'\!-\!p/2}
c_{k'\!+\!p/2}c_{k\!-\!p\!-\!q/2}.
\ea
We are confronted with a generally intractable equation, with all 
possible momentum-preserving processes contributing to the equation of 
motion with a certain matrix element. 
In the RPA theory\cite{pines}, the four-point operator $\cdag\cdag c c$ 
is replaced by the contraction of one of the $\cdag c$ pairs multiplied by 
an operator of the form $\cdag_{k'+q}c_{k'}$. The contraction gives zero 
from translation invariance unless the momentum indices are the same, in 
which case we get $\cdag_{k'}c_{k'}\rightarrow n_{k'}$, the 
electron occupation for a given quantum number $k'$. There are four 
ways of pairing one creation and one annihilation operator in
$\cdag\cdag c c$. 
In a recent numerical study\cite{franco}, it was argued that
the ground state of the edge is in fact not translation-invariant but a 
kind of spontaneously broken symmetry state with charge modulation along
the symmetry direction. For such a system, 
$\langle\cdag_{k}c_{k'}\rangle$ is nonzero even for $k\neq k'$. Detailed
discussion of the symmetry breaking state is not pursued here.

When we put all the terms in the RPA together,

\[
[H_{C},\cdag_{k+q/2}c_{k-q/2}]_{\mbox{RPA}} =
(n_{k\!-\!q/2}\!-\!n_{k\!+\!q/2})
\sum_{k'}\,\,
\left[V_{C}(q,k\!\!-\!\!k')\!-\!V_{C}(k\!\!-\!\!k',q)\right]
c^{\dag}_{k'\!+\!q/2}c_{k'\!-\!q/2} 
\]
\vspace{-0.5cm}
\be
 + c^{\dag}_{k\!+\!q/2}c_{k\!-\!q/2}
\sum_{k'}\,\,\left[V_{C}(0,k\!\!-\!\!k')\!-\!V_{C}(k\!\!-\!\!k',0)\right]
(n_{k'\!+\!q/2}\!-\!n_{k'\!-\!q/2}). 
\label{eq:RPA}
\ee
The commutator with the external potential gives

\be
[H_{i},\cdag_{k+q/2}c_{k-q/2}] =\left[
 V_{i}(k\!+\!q/2)\!-\!V_{i}(k\!-\!q/2)\right]\cdag_{k+q/2}c_{k-q/2}.
\label{eq:external}\ee
The significance of Eqs. (\ref{eq:RPA}) and (\ref{eq:external}) can be 
quite easily seen if we first define the
Hartree-Fock potential and the local energy $E(k)$ as

\ba
 V_{HF}(q,k\!-\!k')&=&V_{C}(q,k\!-\!k')-V_{C}(k\!-\!k',q),
 \label{eq:HFpotential} \\
 E(k)&=&V_{i}(k)+\sum_{k'}V_{HF}(0,k\!-\!k')n_{k'}.
\ea
The net equation of motion looks very simple:

\ba
[H,\cdag_{k+q/2}c_{k-q/2}]&=&(n_{k\!-\!q/2}\!-\!n_{k\!+\!q/2})
\sum_{k'}\,\,V_{HF}(q,k\!-\!k')
c^{\dag}_{k'\!+\!q/2}c_{k'\!-\!q/2} \nonumber \\
  &+&\left[E(k\!+\!q/2)-E(k\!-\!q/2)\right]\cdag_{k\!+\!q/2}c_{k\!-\!q/2}.
\label{eq:totalEq}\ea
The first line may be interpreted as the local density fluctuation 
centered at $k$ interacting via the Hartree-Fock potential
$V_{HF}(q,k-k')$ with the density fluctuation at another point $k'$. 
The single-particle energy term $E(k)$ turns out to be the convolution 
of the total static energy (external+Coulomb) at $y$ with $e^{-(y-k)^{2}}$. 
The static Coulomb energy also includes the exchange contribution.

For a translationally invariant two-dimensional electron gas, the Hartree and 
the 
Fock terms will be proportional to $1/q$ and $1/|\vec{k}\!-\!\vec{k'}|$ 
respectively. In the basis of the Landau wavefunctions we adopted 
here, the Hartree 
term no longer depends exclusively on the transferred momentum $q$. From the 
defintion of the Hartree-Fock potential in Eq.\ (\ref{eq:HFpotential}), one can 
see that the Hartree and the Fock terms are related by the interchange of 
the two 
arguments in $V_{C}(q,k)$. In our particular representation, the Hartree-Fock 
term may be understood as simply a Hartree potential with reduced strength. 
Examination of Eq.\ (\ref{eq:HFpotential}) shows that $V_{HF}(q,k)$ becomes 
negative for the separation $k<q$. 

Proper treatment of the dynamics in Eq.\ (\ref{eq:totalEq}) requires 
first the understanding of the ground state configuration, $\{n_{k}\}$. 
At zero temperature, the occupation is either zero or one. 
The minimization of the energy functional

\be
     E[\{n_{k}\}]= \sum_{k}V_{i}(k)n_{k}+
    {1\over 2}\sum_{k,k'}V_{HF}(0,k-k')n_{k}n_{k'}
\label{eq:energyF}
\ee
determines the distribution $\{n_{k}\}$. A thorough discussion of the
ground state is quite subtle and will be relegated to the 
next section. If we allowed the $n_k$ to vary 
continuously between 0 and 1, the variation of the above functional
is straightforward, and the
minimum is given by\cite{efros}

\be
   E(k)=V_{i}(k)+\sum_{k'}V_{HF}(0,k-k')n_{k'}=\mbox{constant.}
\ee
The ground state would be given 
by the screened state for which $E(k)$ is uniform everywhere. One can 
argue, as we will  in the next section, 
that by the coarse-graining of the microscopic configuration 
one does have a smooth variation in $n_{k}$ and 
the uniform energy $E(k)$ in the ground state. 
In that case, 
the single-particle term in Eq.\ (\ref{eq:totalEq}) disappears and 
the solution for the particle-hole dynamics is given in the form

\[
\rho_{\alpha}(q) =  \sum_{k}h_{\alpha}(k)(\cdag_{k+q/2}c_{k-q/2}-
       \delta_{q0}n_{k}), 
\]
\vspace{-0.5cm}
\be
[H,\rho_{\alpha}(q)] =  \omega_{\alpha}(q)\rho_{\alpha}(q).
\label{eq:definerho}
\ee
For the smooth and monotone occupation, we can replace 
$n_{k\!-\!q/2}-n_{k\!+\!q/2}$
by its gradient, and $h_{\alpha}(k)$ becomes the analogue of the classical 
eigenfunction.\cite{conti,han,aleiner} The discrete matrix 
equation satisfied by $h_{\alpha}(k)$ is

\be
\sum_{k'}V_{HF}(q,k\!-\!k')[n_{k'\!-\!q/2}-n_{k'\!+\!q/2}]
h_{\alpha}(k')=\omega_{\alpha}(q)h_{\alpha}(k).
\ee
In principle the eigenfunctions also depend on
$q$, but the dependence is
very weak at small $q$.\cite{thesis}
We will therefore omit the $q$ dependence in
our definition of $h_{\alpha}(k)$. Take $h_{\alpha}(k)$ to be the 
eigenfunction in the presence of the Hartree term alone, then using the 
first-order perturbation theory, the relative correction in the phase 
velocity $v_{\alpha}(q)=\omega_{\alpha}(q)/q$ due to the Fock term is

\be
 {\delta v_{\alpha}(q)\over{v_{\alpha}(q)}}=
 -{{\int\!\!\int h_{\alpha}(k)V_{C}(k\!-\!k',q)
         h_{\alpha}(k')dn_{k}dn_{k'}}\over
  {\int\!\!\int h_{\alpha}(k)V_{C}(q,k\!-\!k')h_{\alpha}(k')dn_{k}dn_{k'}}}.
\ee
The Fock term extends over the distance of the magnetic length,
$|k-k'|\le l_{0}$ while the 
Hartree term is not limited by any such cutoff except the width of the 
edge, $a$. From dimensional analysis alone, one can see that 
the above term 
is of order $l_{0}/a$ which is very small for a compressible edge.
The influence of the Fock term on the dynamics is 
consequently small. 

Because of the symmetry of the kernel $V_{HF}(q,k-k')$ under the
interchange $k\!\leftrightarrow\!k'$
and $q\!\rightarrow\! -q$. one can write it out in terms of the complete
set of eigenfunctions $h_{\alpha}(k)$ as\cite{han,courant}

\be
V(q,k\!-\!k')\!={{2\pi^{2}}\over{\nu L}}
\sum_{\alpha=0}^{\infty}v_{\alpha}(q)h_{\alpha}(k)h_{\alpha}(k').
\label{eq:decomp}
\ee
This way of writing out the kernel will prove useful for the subsequent 
analysis.

The time evolution of the single-particle operator $c_{p}$ is given by

\ba
[H,c_{p}]&=&\sum_{kq}V_{C}(q,k\!-\!p\!-\!q/2)
\cdag_{k\!+\!q/2}c_{p\!+\!q}c_{k\!-\!q/2}-V_{i}(p)c_{p} 
\nonumber\\
    &=&-E(p)c_{p}-{2\pi^{2}\over{\nu L}}
   \sum_{\alpha,q\neq 0}v_{\alpha}(q)h_{\alpha}(p\!+\!q/2)
        \rho_{\alpha}(q)c_{p\!+\!q}.
\label{eq:singleEOM}
\ea
The energy term $E(p)$ 
is obtained from contraction of the operators $\cdag cc$.
We have used the Eq. (\ref{eq:decomp}) to expand the potential term in 
terms of the constituent density modes. 
Unlike the time evolution of the density operators, we have kept the  
scattering terms in the commutator, giving 
rise to the coupling between a single-particle operator and a density 
operator $\rho_{\alpha}(q)$. The sum is restricted to $q\neq 0$ to  
guarantee $c_{p+q}$ does not coincide with $c_{p}$. All the 
diagonal contributions are contained in $E(p)$. All the dynamics of the
system within our RPA framework is contained in 
Eqs. (\ref{eq:totalEq}) and (\ref{eq:singleEOM}).

\section{Edge Reconstruction}
For a very long Hall bar with the electrons occupying the bulk
with uniform density, the electric field generated by the charges 
diverges 
logarithmically as one approaches the edge. Quantum-mechanically the 
divergence is cancelled by the exchange potential, with the result that 
the field remains finite and proportional to the logarithm of the width 
$2W$ of the bar. 
The confining potential is required to compensate for this outward 
force if the charge configuration is to remain stable. It is 
easy to estimate what the critical field is if we use the harmonic 
confining potential given by

\be
 V_{i}(k)={1\over2}\alpha k^{2}.
\ee
The Hartree and Fock energies can be calculated for the electron occupation
$n_{k}=\theta(W-|k|)$. The force exerted at position $k$ due to the electrons 
is

\ba
-{{\partial E_{HF}(k)}\over{\partial k}}
&=&{e^{2}\over{2\pi\kappa}}[
  2\ln\left({{1\!+\!x}\over{1\!-\!x}}\right)+
 e^{-W^{2}(1\!+\!x)^{2}/4}K_{0}(W^{2}(1\!+\!x)^{2}/4)\nonumber \\
  & - &
 e^{-W^{2}(1\!-\!x)^{2}/4}K_{0}(W^{2}(1\!-\!x)^{2}/4)].
\ea
The logarithmic part comes from the Hartree interaction. Close to the 
edge $x=k/W\approx\pm 1$, one of the modified Bessel functions diverges
logarithmically and cancels the divergence of the Hartree force, leaving
the field strength at
$(e^{2}/2\pi\kappa)\ln (W^{2}e^{\gamma}/2)$. The stability is 
achieved if the confining field strength at the edge is greater than the 
repulsive force of the Hartree-Fock potential:

\be
 \alpha W\ge {e^{2}\over{2\pi\kappa}}\ln (W^{2}e^{\gamma}/2).
\label{eq:stability}
\ee
For a sufficiently narrow bar or a weak enough confining potential this 
condition is violated, and a certain amount of charges separate away from 
the bulk. This newly formed ``island" experiences a stronger confining 
field, but it cannot merge with the bulk from which it got 
separated because of the repulsive Hartree-Fock force. As a result, the 
center of the new island will sit at some equilibrium position away from 
the bulk. Approximately, the position of the center will be

\be
 W+\delta W={e^{2}\over{2\pi\kappa\alpha}}\ln{
       {{W^{2}e^{\gamma}}\over 2}}.
\ee
Because the condition Eq.\ (\ref{eq:stability}) is violated in 
order for the islands to form, the right hand side is slightly greater 
than $W$. 

The fate of the newly formed island will be subject to the same stability 
criteria as Eq.\ (\ref{eq:stability}) with the difference that, in place of 
the bare confining potential,  we now use the 
effective potential composed of $V_{i}(k)$ and the interaction 
potential with the ``parent" island. If the new island does not satisfy 
the stability criteria, it will further split into smaller islands until 
the stability is obtained. By continuing to lower the bare potential 
parameter $\alpha$, one should expect this ``reconstruction"\cite{chamon} 
process to repeat itself over many times.

\begin{figure}[ht]
\centering
\leavevmode
%\vspace{-.6in}
\epsfxsize=12cm\epsfysize=7cm\epsfbox{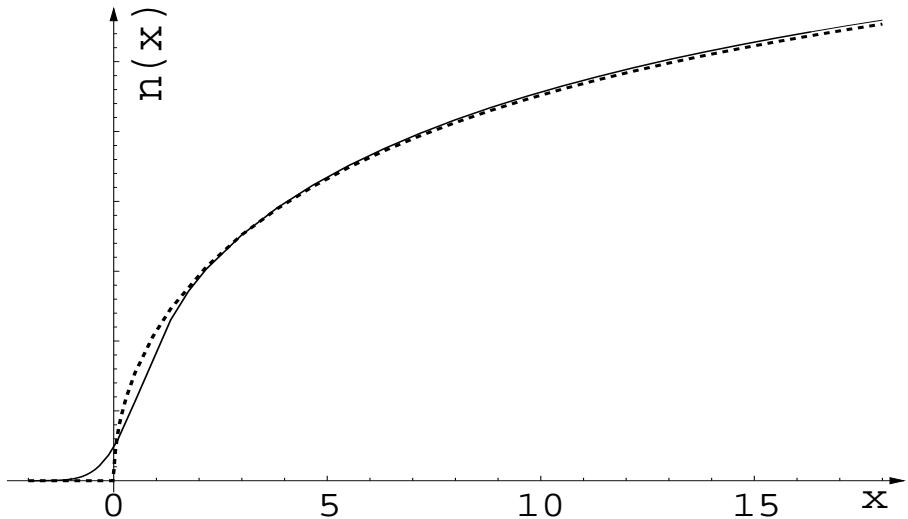}
\caption[Reconstructed Density Plot]
{Plot of the real space density $n(x)=\arctan{(x/a)^{1/2}}$
for $a=10$ (dotted curve). The solid curve is the reconstructed plot
according to Eq. (\ref{eq:mesheq}) with $\Delta=1$. The normalization $A$ has
been chosen to fit the asymptotic values of two curves. Except close
to the edge, the difference is not very noticeable.}
\label{realnkdensity}
\end{figure}

So how is it possible that we recover the classical density profile 
$n(x)$ from an apparently discrete and discontinuous distribution of 
electrons? 
First think of the following correspondence between real and momentum 
space distribution:

\be
  n(x)=A\sum_{k=i\Delta}n_{k}e^{-(x-k)^{2}}.
\label{eq:mesheq}
\ee
The momentum space distribution function is taken to be the value of the 
known real space density $n(x)$ (hence continuous) at a given site,
$n_{k}=n(x=i\Delta)$ ($i$=integer), where
$\Delta$ is the sampling step. The overall normalization on the 
r.h.s. is fixed by $A$. 
Take, for example, the density used by Aleiner and Glazman\cite{aleiner}; 
$n(x)=\arctan{\sqrt{x/a}}$. Can we pick
a mesh size $\Delta$ so that the sum in Eq.\ (\ref{eq:mesheq})
reproduces the real space density?
Considering that the gaussian has a width of unity, it is expected
that $\Delta$ ought to be comparable or less than unity. As it turns out,
using $\Delta=1$ already gives an adequate reproduction of $n(x)$ over an
entire region of space except very close to the edge
(Fig.~\ref{realnkdensity}).
Because of the finite extent of the gaussian, the density does not cut off
where the classical density normally does.

\begin{figure}[ht]
\centering
\leavevmode
%\vspace{-.6in}
\epsfxsize=13cm\epsfysize=9cm\epsfbox{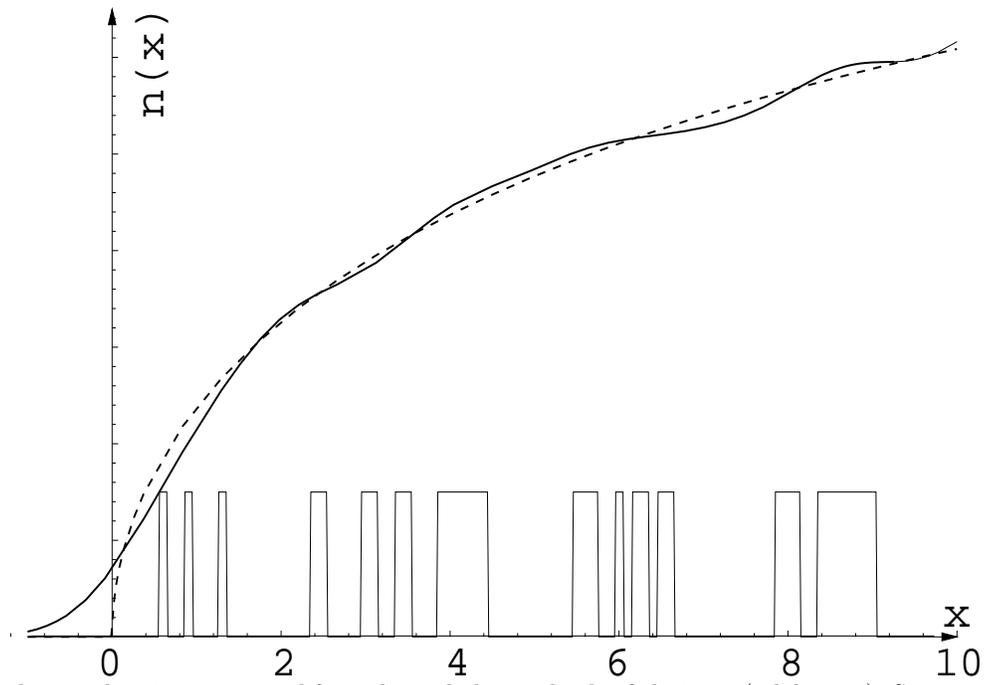}
\caption[Reconstructed Density from the Underlying Islands]
{The real space density constructed from the
underlying islands of electrons (solid curve).
Steps indicate filled states with $\delta x=0.1$ as the width of a
single state. Introducing a finer
step width will smooth out fluctuations around the classical density
(dotted curve) even more.}
\label{discretedensity}
\end{figure}

The spacing between electron states is $2\pi/L$. For a Hall bar of length 
1mm, one magnetic length width is large enough to contain

\be
      N_{c}={L\over{2\pi l_{0}}}\approx\mbox{10}^{4}
\ee
states. 
The microscopic ground state for a very smooth confining potential
is one in which there are a large number of islands, of varying sizes and 
varying separations between the neighbors, 
with the center of each island positioned at 
the local self-consistent potential minimum. Provided the reconstruction 
has reduced the size of the individual island to a sufficiently small 
fraction of the magnetic length,
one expects the coarse graining of occupation to give rise to
the smooth average occupation, which in turn closely follows 
the classical density profile. An illustration of this point is given in 
Fig.~\ref{discretedensity}. Here we have chosen a configuration of 
islands that approximate the real-space density of the electrostatic 
solution, $n(x)=\arctan{(x/a)^{1/2}}$. The underlying occupation $n_{k}$ 
is non-monotonic, but the density itself is a monotone increasing 
function. As figure~\ref{realnkdensity} suggests, the averaged occupation 
$n_{k}$ will follow the density $n(x)$ and hence also monotone. 
One can go back to Eq.\ (\ref{eq:energyF}) and 
treat $n_{k}$ as an averaged, continuous quantity. 
The screened state of the classical electrostatics follows as a 
consequence.

The choice of the island configuration for figure~\ref{discretedensity} 
is arbitrary, and 
does not represent the energy-minimizing state. A detailed numerical study 
in the future is needed to understand the island distribution 
from the proper energetic consideration. The arguments of the preceding
paragraph suggests, we believe, a plausible scenario which also attempts
to link the Hartree-Fock theory (reconstruction picture) with the 
classical picture of monotone density.

The dynamical correspondence between the Hartree-Fock and the classical 
theories can be similarly established. A given island is subject to two 
types of excitations. First is the harmonic motion of its center of mass 
about the local minimum. Alternatively it can 
create a density fluctuation at either edges. As Chamon and 
Wen\cite{chamon} showed, the latter type of excitation does not 
propagate very far, because even a small amount of impurities (backscattering) 
in the substrate will be sufficient to localize the excitation. 
For $N-1$ islands present, there are thus $N$ dynamical degrees of freedom, 
since we have to include the density fluctuations at the boundary of the 
parent island as well as the center coordinate of each island. 
A coherent motion of all the center-of-mass coordinates
would correspond to the EMP mode. All other modes will become the $N-1$ 
acoustic modes, where some of the islands move out of phase with the 
rest.\cite{franco,private} 
In the classical theory, there was no limit on the number of acoustic 
modes present. The present microscopic picture suggests that the upper limit
on the possible modes is the number of islands that make up the edge. 
Further effects such as damping will reduce the number of surviving 
modes. 

The argument presented in this section is valid for integer edge states 
without the extra correlation energy associated with the fractional bulk.
For fractional edge states, Hartree-Fock theory of composite fermions 
had been considered.\cite{brey} The situation is more complicated with 
composite fermions because, in addition to the interplay of the confining 
and the Coulomb potentials, one also has to put the correlation 
energy into play. For integer states, an arbitrary smooth confining 
potential will lead to an arbitrary small island. In the fractional case, 
the correlation energy may overcome the Coulomb repulsion and
resist further splitting even for a very weak confinement. The size
of such an island may not be small enough to validly apply the 
coarse-graining scheme. Consideration of such cases is clearly outside 
the scope of the present paper. We will treat the edge as a set of 
microscopically large but macroscopically small islands whose 
coarse-graining will reproduce the classical picture of screening. It 
should also be noted that Franco's calculation\cite{franco} actually 
yields the screeened state as a ground state without any 
coarse-graining. A thermal averaging of low-lying energy configurations 
will also yield a picture similar to the coarse-graining.

\section{Green's Functions}
With this background, it is now possible to calculate the Green's functions.
Define the one- and two-particle functions as

\[
G_{p}(t) =-i\langle Tc_{p}(t)\cdag_{p}(0)\rangle, 
\]
\vspace{-0.7cm}
\be
B_{\alpha}(q,t;p+q,t')=
\langle T\rho_{\alpha}(q,t)c_{p\!+\!q}(t')\cdag_{p}(0)\rangle.
\label{eq:onetwogreen}
\ee
with $\rho_{\alpha}(q)$ defined as in Eq.\ (\ref{eq:definerho}).
The average $\langle\cdots\rangle$ is over all many-body states
with the Boltzmann weight $e^{-\beta E}$ for each state.
After a straightforward algebra, one has

\[
{{\partial G_{p}(t)}\over{\partial t}}=-i\delta(t)\!-\!
{{2\pi^{2}}\over{\nu L}}\sum_{\alpha,q}v_{\alpha}(q)
h_{\alpha}(p\!+\!q/2)B_{\alpha}(q,t;p\!+\!q,t^{-}), 
\]
\vspace{-0.5cm}
\be
\left\{{\partial\over{\partial t}}\!+\!i\omega_{\alpha}(q)\right\}
B_{\alpha}(q,t;p\!+\!q,t')=ih_{\alpha}(p\!+\!q/2)
\left\{\delta(t)G_{p+q}(t')-\delta(t-t')G_{p}(t)\right\}.
\label{eq:greenEq}\ee
In arriving at the above result, we have used Eqs. (\ref{eq:totalEq}) 
and (\ref{eq:singleEOM}). The sum over the density modes in the first
half of the Eq.\ (\ref{eq:greenEq}) is a result of rewriting the potential
in terms of the eigenfunctions as in Eq. (\ref{eq:decomp}).
The single-particle energy $E(p)$ is taken to be uniform and set to zero  
for all $p$ after coarse-graining. We have used the full equation of 
motion for $c_{p}$, and the approximate RPA equation for the density. In 
effect, RPA truncates the Green's function hierarchy at the two-particle 
level and we have a closed set of equations to solve. The 
symbol $t^{-}$ indicates that 
the time for the operator $c_{p+q}$ in the Eq.\ 
(\ref{eq:onetwogreen}) is supposed to be taken infinitesimally earlier 
than for $\rho_{\alpha}(q)$.

One can remove $B_{\alpha}(q,t;p\!+\!q,t')$ from the above equations. We
adopt the Matsubara representation and write
   
\be
i\omega G_{p}(i\omega)\!=\!-1\!+\!{{2\pi^{2}}\over{\beta\nu L}}
\sum_{\alpha q\Omega}v_{\alpha}(q) h^{2}_{\alpha}(p\!+\!q/2)
{{G_{p\!+\!q}(i\omega\!+\!i\Omega)
\!-\!G_{p}(i\omega)}\over{i\Omega\!-\!\omega_{\alpha}(q)}}.  
\ee
The sum over $q$ and $\Omega$ runs from minus to plus infinity as 
multiples of $2\pi/L$ and $2\pi/\beta$ respectively. The argument 
$\omega$ is an odd multiple of $\pi/\beta$ since it enters in the 
fermionic Green's function. It then follows that $\Omega$ should be an 
even multiple of $\pi/\beta$. 
The thermal sum $\sum_{\Omega}$ can be evaluated with standard
techniques\cite{mahan} 

\ba
{1\over\beta}\sum_{\Omega}{{G_{p+q}(i\omega+i\Omega)}\over
 {i\Omega-\omega_{\alpha}}}&=&
\int_{-\infty}^{\infty}{dx\over{1\!+\!e^{\beta x}}}
{{A_{p+q}(x)}\over{x\!-\!i\omega\!-\!\omega_{\alpha}}}-
 B(\omega_{\alpha})G_{p+q}(i\omega+\omega_{\alpha}), \nonumber \\
{1\over\beta}\sum_{\Omega}{1\over{i\Omega\!-\!\omega_{\alpha}}}&=&
{{-1/2}\over{\tanh{\beta\omega_{\alpha}/2}}}. 
\ea 
We have introduced the
Bose distribution function $B(x)\equiv(e^{\beta x}-1)^{-1}$ above. The
Fermi function will be similarly defined as $F(x)=(e^{\beta x}+1)^{-1}$. 

One can now revert to real frequencies
$i\omega\!\rightarrow\!\omega\!\pm\! i0$ and take the difference
$G_{p}(\omega\!+\!i0)\!-\!G_{p}(\omega\!-\!i0)\equiv -2\pi
iA_{p}(\omega)$, which is the spectral density. The spectral density obeys
the equation

\ba
\omega A_{p}(\omega)&=&-{{2\pi^{2}}\over{\nu L}}\sum_{\alpha q}
v_{\alpha}(q)h^{2}_{\alpha}(p\!+\!q/2)
\left\{F(\omega\!+\!\omega_{\alpha})\!+\!B(\omega_{\alpha})\right\}
A_{p\!+\!q}(\omega\!+\!\omega_{\alpha}) \nonumber \\
 & &+{\pi^{2}\over{\nu L}}\sum_{\alpha q}
v_{\alpha}(q)h^{2}_{\alpha}(p\!+\!q/2)
{{A_{p}(\omega)}\over{\tanh{\beta\omega_{\alpha}/2}}}.
\label{eq:spectral}
\ea
The above equation is rather complicated, and it is hard to believe one can
solve for the spectral density in this form. The reason for the 
difficulty is that Eq.\ (\ref{eq:spectral}) couples the spectral 
densities at different $p$'s. A great simplification can be achieved if 
we assume that we can ignore such a coupling: 

\ba
     A_{p\!+\!q}(\omega)&\rightarrow&A_{p}(\omega), \nonumber \\
     h_{\alpha}(p\!+\!q/2)&\rightarrow&h_{\alpha}(p).
\ea
With this simplification the last term in Eq.\ (\ref{eq:spectral}) 
disappears from symmetry, and we have the following simplified equation 
for $A_{p}(\omega)$:

\be
\omega A_{p}(\omega)=-{{2\pi^{2}}\over{\nu L}}\sum_{\alpha q}
v_{\alpha}(q)h^{2}_{\alpha}(p)
\left\{F(\omega\!+\!\omega_{\alpha})\!+\!B(\omega_{\alpha})\right\}
A_{p}(\omega\!+\!\omega_{\alpha}).
\label{eq:simplespec}
\ee
We have not attempted to justify the above simplification except on the 
ground that we are keeping the lowest order terms in what may be 
considered an expansion in small $q$.

Consider first the zero temperature limit of the above result.
The distribution functions reduce to step functions; $F(x)\!=\!\theta(-x), 
B(x)\!=\!-\theta(-x)$.
The sum of Fermi and Bose functions survives only in the range
$-\omega\!\!<\!\!\omega_{\alpha}\!\!<\!\!0$,
and Eq.\ (\ref{eq:simplespec}) turns
into

\be
\omega A_{p}(\omega)\!=\!{{2\pi^{2}}\over{\nu L}}
\sum_{q}\sum_{\alpha}
v_{\alpha}(q)h^{2}_{\alpha}(p)A_{p}(\omega\!+\!\omega_{\alpha}).
\label{eq:conti}
\ee
The sum over $q$ is restricted to the range 
$-\omega<\omega_{\alpha}(q)<0$ for each $\alpha$. The $q$-dependence of 
the velocity $v_{\alpha}(q)$ is logarithmic if it is the EMP mode, and 
nearly constant for the acoustic ones. One can fix the $q$ entering in 
the velocities with some momentum or the (inverse) length scale 
associated with the 
particular problem at hand. Since the dependence is logarithmic, it 
should not matter which scale we choose to fix $q$. Treating all
$v_{\alpha}$'s as constants, we recover the relation first derived by 
Conti and Vignale\cite{conti}

\be
\omega A_{p}(\omega)=\sum_{\alpha}{\pi\over\nu}h_{\alpha}^{2}(p)\times
\int_{0}^{\omega}A_{p}(\omega')d\omega'.
\ee
The solution is a power-law $A_{p}(\omega)\propto\omega^{\kappa_{p}-1}$ 
with the exponent

\be
  \kappa_{p}=\sum_{\alpha}{\pi\over\nu}h_{\alpha}^{2}(p).
\ee
This is the same tunneling exponent derived previously\cite{conti,han} 
and is given by the sum of the exponents 
associated with each density mode. 
The Green's function approach, rather than contradicting any of the 
previous theories, serves to clarify what physical and mathematical 
approximations were involved in previous theories.

At a non-zero temperature, we follow the method explored by
Everts and Schulz\cite{everts} and write
$A_{p}(\omega)=F(\omega)^{-1}A'_{p}(\omega)$:

\be
\omega A'_{p}(\omega)=
\kappa_{p}\int_{-\infty}^{\infty}d\omega
B(\omega)A'_{p}(\Omega\!-\!\omega).
\label{eq:aprimeeq}
\ee
It requires a series of mathematical
manipulations to solve this equation, and details can be found in the 
appendix. 
We will proceed to the final result for the spectral density:

\be
A_{p}(\omega)=T^{\kappa_{p}\!-\!1}\cosh\left({{\beta\omega}\over2}\right)
\Gamma\!\left(
{\kappa_{p}\over 2}\!+\!{{i\beta\omega}\over{2\pi}}\right)
\Gamma\!\left(
{\kappa_{p}\over 2}\!-\!{{i\beta\omega}\over{2\pi}}\right).
\label{eq:spectralD}
\ee
It is a position-dependent spectral density since $\kappa_{p}$ varies 
with $p$. The tunneling exponent is significantly larger for a compressible
edge with a large number of modes, since $\kappa_{p}$ is 
proportional to the 
squares of the eigenfunctions. For a sharp edge, there are no 
reconstructed islands, and it is only the edge of the bulk that 
fluctuates. The corresponding eigenfunction is a 
structureless constant which, after normalization, becomes 
$1/\sqrt{\pi}$. When sustituted, $\kappa$ equals $1/\nu$ 
independent of $p$. 
This is precisely the result one obtains from bosonization. 

For principal fractions ($\nu=1/m$), the spectral densities can be 
calculated in closed forms. Writing $A_{m}(\omega)$ as the 
spectral density of 
the fraction $1/m$, one can deduce from Eq.\ (\ref{eq:spectralD}) the 
recursion relation

\be
 A_{m+2}(\omega)=[\omega^{2}+(\pi mT)^{2}]A_{m}(\omega)
\label{eq:recursion}
\ee
up to the proportionality factor. Since  $A_{1}(\omega)\!=\!1, 
A_{2}(\omega)\!=\!\omega/\tanh(\beta\omega/2)$, 
we can calculate the spectral 
densities associated with arbitrary fractions $1/m$. Generally, the 
spectral density has an asymptotic limit $\omega^{m-1}$ and 
$T^{m-1}$ for high and low ratios of $\omega/T$ respectively. The even 
denominators have the $\coth(\beta\omega/2)$ factor associated in their 
spectral densities while it is a polynomial for the odd denominators.

\section{Tunneling Currents}
One can imagine the edges of the two quantum Hall liquids brought in
sufficient proximity to allow tunneling between point $p$ on one edge and
$p'$ on the other. The impurity charge is assumed to mediate the 
scattering between edges. Also by locally changing the gate voltage, one 
can induce a finite curvature in the edge shape which results in the loss of
translation symmetry and therefore, a finite amplitude for tunneling.
We will consider a single scattering process 
(Fermi's Golden Rule) with a fixed, constant scattering matrix element. 
The electron tunneling current between the edges is   
proportional to the integral 

\be
\int_{-\infty}^{\infty}
A_{\kappa_{p}}(\omega\!-\!V/2)A_{\kappa_{p'}}(\omega\!+\!V/2)\!\times\!
\{F(\omega\!+\!V/2)\!-\!F(\omega\!-\!V/2)\}d\omega.
\ee
where $V$ is the chemical potential difference of
the two edges. For the tunneling current at a finite temperature, one needs
to use the expression of the spectral density in
Eq.\ (\ref{eq:spectralD}). The tunneling current 
characterized by two tunneling exponents $(\kappa_{p},\kappa_{p'})$ turns 
out to be ($\kappa=\kappa_{p}+\kappa_{p'}$)

\be
I_{\kappa}(V)=T^{\kappa-1}\sinh\left({{\beta V}\over2}\right)
\Gamma\left({\kappa\over 2}+i{{\beta V}\over{2\pi}}\right) 
\Gamma\left({\kappa\over 2}-i{{\beta V}\over{2\pi}}\right).
\label{eq:tunnelI}
\ee
As it happens, the tunneling current is only dependent on the sum of the
tunneling exponents. 
Compare this expression with that for the spectral density, Eq.\ 
(\ref{eq:spectralD}), and one finds they are very similar. The recursion 
relation for the current will be the same as derived in Eq.\ 
(\ref{eq:recursion}). We need to know the $I_{1}(V)$ and $I_{2}(V)$
to derive the tunneling current for all integer values of $\kappa$. 

\be
   I_{1}(V)=\tanh\left({{\beta V}\over 2}\right),\,\,
   I_{2}(V)=V.
\ee
We can specialize to the limit of a sharp edge, where the significance
$\kappa$ is the sum of the two inverse fractions for each edge;
$\kappa=m+n$. 
The tunneling formula in this case has been derived by Wen\cite{wen} and 
agrees with our result.
One sees that, for example, a tunneling between $\nu=1$ and $\nu=1/3$
edge ($\kappa=4$) would give the same $I(V)$ curve as tunneling between
two 1/2 edges. The same is true of two 1/3 edges and 1 and 1/5. This 
symmetry is inherent in the chiral Luttinger liquid action\cite{chamon96}
and is explicitly demonstrated in the above calculation. In the zero 
temperature limit, $I_{\kappa}$ is proportional to $V^{\kappa-1}$. 
For a high enough temperature 
we are back in the ohmic regime with the conductance proportional to 
$T^{\kappa-2}$. The crossover between the two regimes occurs around 
$T/V\equiv 1$.

\section{Conclusion}
We have demonstrated that all the electronic spectral
properties of
the quantum Hall edges, both wide and sharp, can be derived by using the
equation of motion method and the conventional many-body theory.
We relate the bosonization theory of the edge state to the random phase
approximation and argue that they are in fact equivalent. It may happen 
that the real system is subject to a rather large dissipation, or that the
terms neglected in the random phase approximation actually contribute 
significantly. These possibilities and their consequences on tunneling are 
not understood in detail. One can, however, rather easily see the 
consequence of a stable density mode whose dispersion is not linear with
$q$ but goes as $q^{\eta}$ where $\eta$ may be an arbitrary positive 
number. Going back to Eq.\ (\ref{eq:conti}), it can be shown that a 
single-mode edge at zero temperature with the tunneling exponent 
$\kappa$ and the collective dispersion exponent $\eta$ obeys

\be
\omega A(\omega)={\kappa\over\eta}\int_{0}^{\omega}A(\omega')d\omega'.
\label{eq:nonlineardisp}
\ee
The actual tunneling exponent is $\kappa/\eta$. The suppression of the 
low-lying excitations leads to the {\it enhanced} 
tunneling at low energy. It is therefore crucial in the
determination of the tunneling exponent that we know the precise power-law
dependence of the dispersion as well.
We have been somewhat cavalier about taking the results of
our calculation to both the odd and even fractions, while there is a 
marked difference in the bulk properties of the even and odd denominator 
states. It is 
still true there are gapless modes close to the edge for any fractions. 
For the gapless, even-denominator states the edge excitation will couple
to the bulk fluctuation, possibly changing the dispersion relation to deviate
from linearity. Our formula in Eq.\ (\ref{eq:nonlineardisp})
suggests that the spectral density will behave as
$\omega^{m/\eta-1}$ for a filling 
fraction $1/m$ whose gapless mode has a dispersion like 
$\omega(q)\propto q^{\eta}$. Presence of several gapless modes will
contribute independently to the exponent in the usual manner.

We have used the electronic basis to expand the second-quantized 
Hamiltonian and consequently, we are not able to discuss the 
fractionally-charged quasiparticles within the current framework. 
Tunneling experiments between the chiral edges through the fractional 
bulk\cite{milliken96} is outside the scope of the theory presented in 
this paper. For this purpose, one must still rely on the effective
theories.\cite{han,theory} For the recent experiment of {\it electron} 
tunneling 
to the edge of the fractional liquid,\cite{chang} our result gives a 
good fit to the observed I-V curve.

The author wishes to acknowledge helpful conversations  
with Albert Chang, Michael Geller, Matt Grayson, David Thouless and 
Carlos Wexler. 
This work was supported by an NSF grant, DMR-9628345.

\appendix\section*{Evaluating the Spectral Density at a Finite Temperature}  

One wants to solve the Eq.\ (\ref{eq:aprimeeq}),

\ba
\omega A'_{p}(\omega)&=&
\kappa_{p}\int_{-\infty}^{\infty}d\omega
B(\omega)A'_{p}(\Omega\!-\!\omega), \nonumber \\
-i{{dA'_{p}(t)}\over{dt}}&=&\kappa_{p}B(t)A'_{p}(t).
\ea
We have written the Fourier-transformed version of the equation in the 
second line. The solution to this first-order differential equation 
is

\be
 A'_{p}(t)=A'_{p}(0)\exp[i\kappa_{p}\int_{0}^{t}B(t')dt'].
\ee
The Fourier transform of the Bose function is

\be
B(t)=\int_{-\infty}^{\infty}
e^{i\omega t-\gamma|\omega|}B(\omega)d\omega=
-i{d\over{dt}}\ln\Gamma\left({{\gamma\!+\!it}\over\beta}\right)
                 \Gamma\left(1\!+\!{{\gamma\!-\!it}\over\beta}\right).   
\ee
We have introduced the convergence factor $\gamma$ in the definition of
$B(t)$. Substituting (A.3) to (A.2) and Fourier-transforming back, 
the final result for $A_{p}(\omega)$ is 
(ignoring the overall constant)

\be
A_{p}(\omega)=\cosh\left({{\beta\omega}\over2}\right)B^{-\kappa_{p}}
     \left({\gamma\over\beta},
             1\!+\!{\gamma\over\beta}\right)
\int_{-\infty}^{\infty}{{dt}\over{2\pi}}e^{-i\omega t}
B^{\kappa_{p}}\left({1\over2}\!+\!
{{\gamma\!+\!it}\over\beta},{1\over2}\!+\!{{\gamma\!-\!it}\over\beta}\right).
\ee
In this form, the integrand is free of singularities, and one can take 
$\gamma$ to be zero. The beta function $B(x,y)$
in front of the integral gives 
$T^{\kappa_{p}}$ and up to a constant, the beta function inside the integral
becomes $1/\cosh^{\kappa_{p}}(\pi t/\beta)$ and one can carry out the
integration to yield the Eq.\ (\ref{eq:spectralD}).

\end{document}